\documentclass{article}
\usepackage{spconf,amsmath,amssymb,graphicx}
\usepackage{booktabs}
\usepackage{multirow}
\usepackage{url}

\newcommand{\zs}{\textsc{zs}}

\title{Transductive Zero-Shot Audio Classification\\ with Audio--Language Models}

\name{Jingwen Zhou \qquad Mingzhe Wang}
\address{Xidian University, Xi'an, China}

\begin{document}
\ninept
\maketitle

\begin{abstract}
Contrastive language--audio pretraining (CLAP) enables zero-shot audio classification, but standard inference classifies each clip in isolation and ignores the structure of the unlabeled test set. We present the first systematic study of TransCLIP-style transductive inference for CLAP: a text-anchored spherical Gaussian-mixture EM that refines zero-shot posteriors using the audio-embedding statistics of the test batch, with no labels, no gradients, and negligible compute (about 15\,ms on one CPU core for 2{,}000 clips). Across ESC-50, UrbanSound8K, and VocalSound, this consistently improves top-1 accuracy by $+4.6$ to $+9.2$ points over the zero-shot baseline (e.g., $89.1\!\rightarrow\!94.8\%$ on ESC-50, $73.8\!\rightarrow\!81.8\%$ on UrbanSound8K). We further show that the gain (i) is governed by a simple operating boundary---roughly $2.5$ test samples per class per batch are required, with diminishing returns beyond ${\sim}5$; (ii) is complementary to entropy-guided prompt weighting, with the combination reaching $96.2\%$ on ESC-50; and (iii) attenuates but remains positive under long-tailed batches ($+4.9\!\rightarrow\!+3.1$ points at a $20{:}1$ imbalance), which we report as an explicit limitation. We also document a negative result: on TUT Urban Acoustic Scenes 2018, where zero-shot CLAP is near chance, transduction has no signal to amplify.
\end{abstract}


\section{Introduction}
\label{sec:intro}

Audio--language models such as CLAP \cite{wu2023clap,elizalde2023clap,elizalde2024msclap} transfer the contrastive recipe of CLIP \cite{radford2021clip} to audio: a dual encoder---typically a transformer audio tower \cite{chen2022htsat} pretrained at AudioSet scale \cite{gemmeke2017audioset} paired with a language model---aligns audio clips with natural-language captions. Any label set can then be classified zero-shot by embedding prompts such as \emph{``the sound of a dog''} and selecting the nearest text embedding, generalizing earlier CLIP-distilled audio models \cite{guzhov2022audioclip,wu2022wav2clip} and pre-ALM zero-shot audio classification via semantic label embeddings \cite{xie2021zeroshot}. Zero-shot CLAP is now a standard baseline for environmental-sound tasks, reaching 82--91\% top-1 accuracy on ESC-50 depending on prompts and checkpoints \cite{wu2023clap}. Its accuracy, however, is notoriously prompt-sensitive: gains have been pursued by describing sounds with LLM-generated captions \cite{ghosh2024reclap}, learning prompts from few labeled shots \cite{hanif2024palm}, or hand-crafting template ensembles, mirroring the prompt-engineering line in vision \cite{zhou2022coop,zhou2022cocoop}.

Standard zero-shot inference is \emph{inductive}: every test clip is classified independently. In practice, test clips rarely arrive alone---they come as files in a folder, segments of a recording session, or items in an evaluation fold. The unlabeled batch itself carries information: clips of the same class cluster in the audio embedding space, and that structure can correct individual errors of the audio--text alignment. Exploiting unlabeled test structure is the classical idea of transduction \cite{vapnik1998statistical}, extensively validated in few-shot image classification \cite{liu2019tpn,ziko2020laplacian}. For vision--language models, transductive zero-shot inference was recently revisited \cite{martin2024transductive}, and TransCLIP \cite{zanella2024transclip} formalized it for CLIP at scale, showing that a Gaussian-mixture objective with a text-prior penalty, optimized over the unlabeled test set, boosts zero-shot accuracy at negligible cost. The related test-time adaptation literature \cite{liang2024ttasurvey}---e.g., TENT \cite{wang2021tent} and test-time prompt tuning \cite{shu2022tpt}---pursues a similar goal but requires gradient updates of parameters or prompts. In the audio modality, recent work adapts CLAP at test time on the \emph{text} side, e.g., entropy-guided prompt weighting \cite{entropyprompt2026}, echoing zero-shot prompt-weighting ideas from vision \cite{allingham2023prompt}, but the structure of the unlabeled \emph{audio} batch remains unexploited.

This paper asks: \emph{does transductive inference transfer from vision--language to audio--language models, and under which operating conditions?} We adapt the core of TransCLIP to CLAP in a deliberately minimal form---a text-anchored spherical GMM-EM over the test batch (Sec.~\ref{sec:method})---and run, to our knowledge, the first systematic study across four audio benchmarks. Our contributions:
\begin{itemize}
\itemsep0.1em
\item \textbf{Consistent gains.} Label-free, gradient-free transductive refinement improves CLAP zero-shot top-1 accuracy by $+4.6$ to $+9.2$ points on ESC-50, UrbanSound8K, and VocalSound under official protocols (Sec.~\ref{ssec:main}).
\item \textbf{An operating boundary.} A $5{\times}3$ grid over batch size and label-space size on ESC-50 shows the gain is governed by the \emph{samples-per-class-per-batch} ratio $N/C$: positive above ${\sim}2.5$, diminishing returns beyond ${\sim}5$, negative below ${\sim}1.5$ (Sec.~\ref{ssec:grid}).
\item \textbf{Complementarity with prompt-side adaptation.} Under an identical protocol, audio-side transduction yields larger gains than a batch-level variant of entropy-guided prompt weighting \cite{entropyprompt2026}, and the two compose to $96.2\%$ on ESC-50 (Sec.~\ref{ssec:entropy}).
\item \textbf{Honest failure modes.} Gains attenuate (but stay positive) under long-tailed batches, and vanish on acoustic scenes where zero-shot CLAP is near chance (Secs.~\ref{ssec:longtail}, \ref{ssec:tut}).
\end{itemize}

\section{Method}
\label{sec:method}

\subsection{Zero-shot CLAP inference}
Let $f_a$ and $f_t$ denote the CLAP audio and text encoders. For a test batch $\{x_i\}_{i=1}^{N}$ and $C$ classes, we compute $\ell_2$-normalized audio embeddings $\mathbf{a}_i$ and text embeddings $\mathbf{t}_c$, obtained by encoding one prompt (\emph{single}) or averaging several templates before re-normalization (\emph{ensemble}). Zero-shot posteriors are
\begin{equation}
z^{0}_{ic}=\frac{\exp\!\left(\tau\,\mathbf{a}_i^{\top}\mathbf{t}_c\right)}{\sum_{c'}\exp\!\left(\tau\,\mathbf{a}_i^{\top}\mathbf{t}_{c'}\right)},
\label{eq:zs}
\end{equation}
where $\tau$ is the learned logit scale of the checkpoint ($\tau{=}18.66$ for \texttt{laion/clap-htsat-unfused}). Inductive zero-shot prediction is $\arg\max_c z^0_{ic}$.

\subsection{Text-anchored transductive GMM-EM}
Following the transductive view of TransCLIP \cite{zanella2024transclip}, we model the batch as a mixture of $C$ directional components on the unit sphere \cite{banerjee2005vmf}, one per class, with the \emph{text embeddings acting as anchors}, and fit it by EM \cite{dempster1977em}. We keep only the essential ingredients of \cite{zanella2024transclip}---no covariance estimation, no Laplacian term---so the procedure is three lines of linear algebra. Initializing $z\!\leftarrow\!z^{0}$, each iteration performs
\begin{align}
\textbf{M-step:}\quad & \boldsymbol{\mu}_c=\frac{\sum_i z_{ic}\,\mathbf{a}_i}{\bigl\lVert\sum_i z_{ic}\,\mathbf{a}_i\bigr\rVert},
\label{eq:mstep}\\
\textbf{E-step:}\quad & z_{ic}\propto\exp\Bigl(\beta\,\tau\,\mathbf{a}_i^{\top}\boldsymbol{\mu}_c+(1-\beta)\log z^{0}_{ic}\Bigr),
\label{eq:estep}
\end{align}
where $\beta\in[0,1]$ balances the batch-estimated audio likelihood against the frozen text prior; the $\log z^0$ term is the text-prior KL penalty of \cite{zanella2024transclip} in closed form. We use $\beta{=}0.5$ and a \emph{fixed budget of 3 iterations with no early stopping or convergence criterion}, identical across all datasets; no dataset-specific tuning is performed. Accuracy typically plateaus by the third iteration at sufficient batch size, while at very small $N/C$ additional iterations can degrade it (Sec.~\ref{ssec:grid}). The cost is $O(NCd)$ per iteration: 3 iterations on $2{,}000$ clips with $C{=}50$, $d{=}512$ take ${\approx}15$\,ms on a single CPU core (measured).

We also evaluated classical graph-based label propagation \cite{zhou2004llgc} on a symmetrized $k$NN graph ($k{=}8$, damping $\alpha{=}0.7$) over the same audio embeddings, as a non-parametric alternative.

\subsection{Entropy-guided prompt weighting (compared baseline)}
As a prompt-side point of comparison we implement a \emph{batch-level simplified variant} of entropy-guided prompt weighting \cite{entropyprompt2026}: given $M$ templates, we compute per-template posteriors on the batch, their mean prediction entropies $H_m$, and weights $w=\operatorname{softmax}(-H/T_w)$ with $T_w{=}0.01$. (The original method optimizes the weights with an iterative, regularized objective; our variant keeps only its core principle, low entropy as a confidence proxy.) Weighted logits give the prompt-adapted zero-shot prediction, and the re-normalized weighted text embedding $\sum_m w_m\mathbf{t}^{(m)}_c$ can serve as the anchor in Eqs.~\eqref{eq:mstep}--\eqref{eq:estep}, composing both methods.

\subsection{Relation to prior adapters}
\label{ssec:relation}
Our procedure occupies a deliberately extreme point in the adaptation design space. Relative to TransCLIP \cite{zanella2024transclip}, we drop the per-class covariances and the Laplacian affinity term and keep only mean estimation under a text prior; this removes all graph construction and makes the update strictly $O(NCd)$. Relative to the Dirichlet-based transductive CLIP of \cite{martin2024transductive}, we operate directly on embeddings rather than on probability features, and require no inner optimization loop. Unlike TENT \cite{wang2021tent} or TPT \cite{shu2022tpt}, no parameter or prompt receives gradient updates, so the method cannot drift and needs no learning rate; unlike prompt learning \cite{zhou2022coop,zhou2022cocoop,hanif2024palm} it uses zero labels. Entropy-based prompt weighting \cite{entropyprompt2026,allingham2023prompt} is complementary rather than competing: it improves the \emph{prior} $z^0$ (text side), whereas Eqs.~\eqref{eq:mstep}--\eqref{eq:estep} improve the \emph{likelihood} (audio side); Sec.~\ref{ssec:entropy} tests both axes and their composition.

\section{Experiments}
\label{sec:exp}

\noindent\textbf{Setup.} We use the public \texttt{laion/clap-htsat-unfused} checkpoint \cite{wu2023clap} (153M parameters) frozen, audio resampled to 48\,kHz. Prompts: \emph{single} = ``the sound of a \{\}''; \emph{ensemble} = mean of 4 templates.\footnote{``the sound of a \{\}'', ``the sound of \{\}'', ``this is a sound of \{\}.'', ``\{\} can be heard''. Class names use spaces for underscores.} Datasets: ESC-50 \cite{piczak2015esc50} (2{,}000 clips, 50 classes, official 5-fold), UrbanSound8K \cite{salamon2014us8k} (8{,}732 clips, 10 classes, official 10-fold), VocalSound \cite{gong2022vocalsound} (test split, 3{,}591 clips, 6 classes), and TUT Urban Acoustic Scenes 2018 \cite{mesaros2018tut} (class-balanced 2{,}000-clip subset, 10 scenes).\footnote{Audio from Hugging Face mirrors \texttt{ashraq/esc50}, \texttt{danavery/urbansound8K}, \texttt{lmms-lab/vocalsound}, \texttt{mteb/tut-acoustic-scenes-mini}; official metadata (filenames, folds, labels) verified. Code, scripts, and precomputed embeddings will be released upon publication.} Transductive batches are always drawn \emph{within} an official fold (or the test split); accuracy is averaged over folds; stochastic batch splits use 3 seeds (std $\le 0.4$ points for all entries of Table~\ref{tab:main}). No labels are used at any point.

\subsection{Main results across datasets}
\label{ssec:main}

\begin{table}[t]
\centering
\caption{Top-1 accuracy (\%): zero-shot CLAP vs.\ transductive GMM-EM (3 iterations) with the transductive batch-size ablation. ``full'' = one batch per official fold / test split (400 for ESC-50, 806--990 for US8K, 3{,}591 for VocalSound). $\Delta$ is EM@full minus zero-shot.}
\label{tab:main}
\small
\setlength{\tabcolsep}{2.0pt}
\begin{tabular}{llccccc}
\toprule
Dataset & Prompt & \zs{} & EM@64 & EM@256 & EM@full & $\Delta$ \\
\midrule
\multirow{2}{*}{ESC-50} & single & 85.15 & 87.05 & 93.05 & \textbf{94.30} & $+9.15$ \\
 & ens. & 89.10 & 88.70 & 93.92 & \textbf{94.75} & $+5.65$ \\
\midrule
\multirow{2}{*}{US8K} & single & 73.83 & 80.04 & 81.13 & \textbf{81.80} & $+7.98$ \\
 & ens. & 73.29 & 79.70 & 80.55 & \textbf{81.10} & $+7.82$ \\
\midrule
\multirow{2}{*}{VocalSound} & single & 65.72 & 73.67 & 73.71 & \textbf{73.99} & $+8.27$ \\
 & ens. & 75.27 & 79.19 & 79.62 & \textbf{79.84} & $+4.57$ \\
\bottomrule
\end{tabular}
\end{table}

\begin{figure}[t]
\centering
\includegraphics[width=\linewidth]{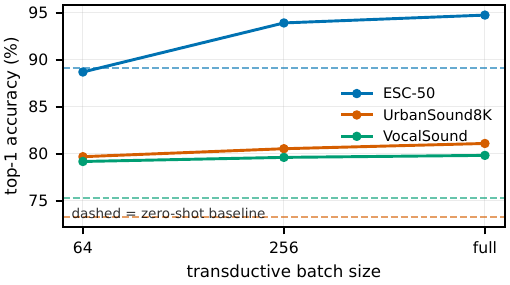}
\caption{Transductive accuracy (ensemble prompts, EM, 3 iterations) vs.\ batch size on the three event/vocal datasets; dashed lines are the corresponding inductive zero-shot baselines. Gains grow monotonically with batch size and are positive everywhere except ESC-50 at batch 64, where $N/C{=}1.3$ (cf.\ Fig.~\ref{fig:grid}).}
\label{fig:batch}
\end{figure}

Table~\ref{tab:main} and Fig.~\ref{fig:batch} report the main comparison. The zero-shot baselines fall in the expected literature range (85.2/89.1\% on ESC-50; 73--74\% on UrbanSound8K). Text-anchored EM improves \emph{every} dataset--prompt configuration, by $+4.6$ to $+9.2$ points at the full-fold batch. The gain grows monotonically with batch size and is largest exactly where the text prior is weakest (single-prompt VocalSound: $65.7\!\rightarrow\!74.0\%$): batch statistics partly substitute for prompt engineering.

\textbf{Iteration dynamics.} The EM refinement is front-loaded but not one-shot: on ESC-50 (ensemble, full fold) accuracy moves $93.70\!\rightarrow\!94.65\!\rightarrow\!94.75\%$ over the three iterations, on UrbanSound8K $79.36\!\rightarrow\!81.10\%$, and on VocalSound $79.39\!\rightarrow\!79.84\%$; i.e., the first iteration captures 70--90\% of the total gain and the trajectory is monotone whenever $N/C$ is sufficient. Below the boundary the trajectory \emph{reverses}: at batch 64 on ESC-50 the ensemble run decays $90.22\!\rightarrow\!89.28\!\rightarrow\!88.70\%$, because noisy means are re-estimated from their own erroneous assignments. A fixed small budget (3 iterations) is therefore not merely a convenience but a guard against this compounding regime; an adaptive stopping rule based on the estimated $N/C$ is an obvious refinement.

\textbf{Label propagation fails where EM succeeds.} $k$NN label propagation is clearly negative at small and medium batches on ESC-50 with ensemble prompts ($89.1\!\rightarrow\!87.5\%$ at batch 64 and $86.9\%$ at batch 256, best iteration) and at best marginal at the full fold ($89.4\%$, $+0.3$; later iterations decline to $88.4\%$). We therefore advocate the text-anchored parametric form rather than graph smoothing in this regime; Sec.~\ref{sec:discussion} quantifies why.

\subsection{Operating boundary: batch size $\times$ label-space size}
\label{ssec:grid}

\begin{figure}[t]
\centering
\includegraphics[width=\linewidth]{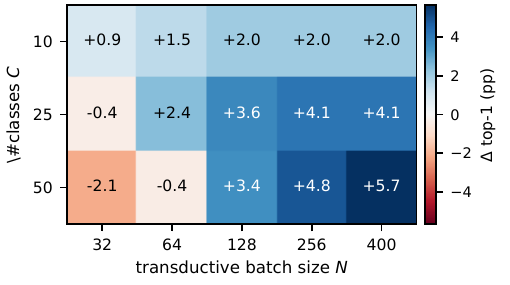}
\caption{Gain of transductive EM over zero-shot (percentage points) on ESC-50 over a grid of batch size $N\in\{32,64,128,256,400\}$ and label-space size $C\in\{10,25,50\}$. Each cell is the mean over 9 runs (3 random class subsets $\times$ 3 batch splits; 3 runs for $C{=}50$, where the class set is fixed); per-cell std is at most 1.2 points and is dominated by the class-subset draw. The gain is governed by $N/C$: positive above ${\sim}2.5$, diminishing returns beyond ${\sim}5$ (plateau around ${\sim}10$), negative below ${\sim}1.5$.}
\label{fig:grid}
\end{figure}

Fig.~\ref{fig:grid} maps the regime where transduction helps. Holding the protocol fixed, we vary the batch size $N$ and subsample the ESC-50 label space to $C$ classes. The pattern is clean: cells with $N/C\gtrsim 2.5$ show positive gains (e.g., $+3.4$ points at $N{=}128$, $C{=}50$), returns diminish beyond $N/C\approx 5$ and plateau around $N/C\approx 10$ (e.g., $C{=}10$: $+2.0$ points from $N{=}128$ on), and cells with $N/C\lesssim 1.5$ degrade (worst: $-2.1$ points at $N{=}32$, $C{=}50$). This yields a one-line deployment rule and explains the batch-64 dip for ESC-50 in Table~\ref{tab:main} ($N/C{=}1.3$).

\subsection{Comparison and composition with entropy-guided prompts}
\label{ssec:entropy}

\begin{table}[t]
\centering
\caption{Audio-side transduction vs.\ text-side entropy-guided prompt weighting (batch-level simplified variant of \cite{entropyprompt2026}; pool of 8 templates; weight temperature $0.01$), top-1 \%. ``unif.''\ = uniform template averaging over the same 8-template pool. Zero-shot columns are batch-independent.}
\label{tab:entropy}
\setlength{\tabcolsep}{2.5pt}
\small
\begin{tabular}{llcccc}
\toprule
Dataset & Batch & \zs{} unif. & \zs{} entropy & EM (unif.) & EM (entropy) \\
\midrule
\multirow{2}{*}{ESC-50} & 256 & 88.00 & 92.00 & 93.92 & \textbf{95.27} \\
 & full & 88.00 & 92.00 & 94.85 & \textbf{96.20} \\
\midrule
\multirow{2}{*}{US8K} & 256 & 72.89 & 76.33 & 80.55 & \textbf{80.86} \\
 & full & 72.89 & 76.33 & \textbf{81.29} & 81.04 \\
\bottomrule
\end{tabular}
\end{table}

Table~\ref{tab:entropy} compares both adaptation axes under an identical protocol with a pool of 8 templates; the original method of \cite{entropyprompt2026} was validated across five datasets including FSD50K \cite{fonseca2022fsd50k}, and our simplified variant reproduces gains of comparable magnitude here. Entropy weighting effectively performs \emph{unsupervised template selection}: at low temperature the weights collapse onto the lowest-entropy template (``this is a sound of \{\}.''), which coincides with the oracle-best single template on ESC-50 ($92.0\%$)---a useful label-free diagnostic in itself. Transduction yields larger gains ($+4.2$ to $+8.4$ points over the respective zero-shot initialization) and operates on the orthogonal, audio-side axis: on ESC-50 the two compose additively at both batch sizes, reaching $\mathbf{96.2\%}$ at the full fold ($+4.2$ on top of entropy weighting, $+1.35$ on top of EM alone) and $95.3\%$ at batch 256. On UrbanSound8K the composition is essentially neutral: $+0.3$ points over EM alone at batch 256, $-0.25$ at the full fold---the prompt-side gain is absorbed once the audio-side statistics are strong enough. Composition was thus either clearly beneficial or nearly neutral in our experiments, never harmful beyond $0.3$ points, which supports treating the two axes as independent knobs.

\subsection{Class-prior sensitivity under long-tailed batches}
\label{ssec:longtail}

\begin{table}[t]
\centering
\caption{Long-tailed transductive batches on ESC-50: per fold, $N{=}200$ clips sampled without replacement under an exponential class prior with the given head-to-tail ratio (random class ranking per seed; 5 seeds; ensemble prompts). Head/tail = upper/lower half of classes by sampled frequency. EM3 = transductive EM, 3 iterations.}
\label{tab:longtail}
\setlength{\tabcolsep}{2.5pt}
\small
\begin{tabular}{lccccc}
\toprule
Imbalance & \zs{} & EM3 & $\Delta$ & Head \zs{}$\rightarrow$EM3 & Tail \zs{}$\rightarrow$EM3 \\
\midrule
$1{:}1$ & 89.54 & 94.46 & $+4.92$ & $91.34\rightarrow 95.71$ & $87.78\rightarrow 93.21$ \\
$5{:}1$ & 89.64 & 93.14 & $+3.50$ & $90.99\rightarrow 94.02$ & $87.26\rightarrow 91.52$ \\
$20{:}1$ & 89.38 & 92.46 & $+3.08$ & $90.72\rightarrow 93.13$ & $85.83\rightarrow 90.68$ \\
\bottomrule
\end{tabular}
\end{table}

Eq.~\eqref{eq:estep} implicitly assumes a roughly uniform class prior within the batch. Table~\ref{tab:longtail} quantifies the violation cost: under a $20{:}1$ exponential prior the gain shrinks by about one third ($+4.9\!\rightarrow\!+3.1$ points) but remains clearly positive at every imbalance level tested. Two finer observations are worth recording. First, the damage is asymmetric in the expected direction---the zero-shot tail accuracy itself drops with imbalance ($87.8\!\rightarrow\!85.8\%$, since tail classes are represented by very few clips per batch)---yet tail classes still receive the largest absolute improvement from transduction ($+4.9$ points at $20{:}1$, vs.\ $+2.4$ for head classes): even one or two same-class clips suffice to pull the text-anchored mean toward the right region, whereas head classes are already near their large-sample estimate. Second, the iteration trajectory remains monotone under imbalance (e.g., $92.0\!\rightarrow\!92.5\%$ from iteration 1 to 3 at $20{:}1$), so the degradation is a bias effect of the implicit uniform prior, not an instability. We report this as a limitation of the uniform-prior form; estimating the batch prior or adding a Sinkhorn-style marginal constraint \cite{cuturi2013sinkhorn} is a natural fix left to future work.

\subsection{Negative result: acoustic scenes}
\label{ssec:tut}
On TUT Urban Acoustic Scenes 2018 \cite{mesaros2018tut}, zero-shot CLAP is near chance ($11.7\%$ ensemble, $10.1\%$ single; chance $=10\%$), and none of 7 additional scene-oriented prompt variants (e.g., ``an audio recording captured in a \{\}'') exceeded $11.2\%$. A supervised linear probe on the same audio embeddings reaches only about $41\%$ on TUT versus about $87\%$ on UrbanSound8K, indicating the audio tower itself barely separates scene categories. Transduction accordingly yields only $+1.4$ points ($11.7\!\rightarrow\!13.1\%$): \emph{transduction amplifies existing alignment signal; it cannot create one}. We include this boundary case deliberately, as it delimits the method's scope to domains where CLAP has non-trivial zero-shot competence.

\section{Discussion}
\label{sec:discussion}

\textbf{Why parametric EM succeeds where graph propagation fails.}
Label propagation and text-anchored EM consume the same unlabeled evidence but aggregate it at different granularities: propagation moves probability mass along \emph{local} $k$NN edges, whereas the M-step \eqref{eq:mstep} pools \emph{all} $N$ posterior-weighted embeddings into $C$ global class means. To quantify the difference, we measured the label purity of the $k{=}8$ cosine-NN graph actually used by LP on ESC-50 (label used for measurement only): within a full fold ($N/C{=}8$), $76.8\%$ of neighbors share the query's class, but within batch-64 subgraphs ($N/C{=}1.3$) purity collapses to $12.0\%$. Crucially, this is information starvation rather than a retrieval failure of the encoder: under hypergeometric sampling a 64-clip batch contains only ${\approx}1.1$ same-class peers per clip on average, capping the attainable 8-NN purity at ${\approx}13.8\%$---the observed $12.0\%$ is near that ceiling. A graph built in this regime is dominated by cross-class edges by construction, so propagation necessarily mixes clusters; iterating makes it worse even at the full fold ($89.40\!\rightarrow\!88.85\!\rightarrow\!88.40\%$), the classic over-smoothing pattern. The parametric form degrades far more gracefully for two reasons: (i) a class mean is a \emph{global} statistic whose error decreases with every same-class clip anywhere in the batch, not just among nearest neighbors; and (ii) the frozen text prior in \eqref{eq:estep} ($\beta{=}0.5$) bounds how far a corrupted mean can drag the posterior, a safeguard label propagation lacks. This also explains the shape of Fig.~\ref{fig:grid}: the mean-estimation error of class $c$ scales with the inverse square root of its batch count, so gains rise steeply between $N/C\!\approx\!1.5$ and $5$ and then saturate once means are accurate.

\textbf{Failure-mode taxonomy.} Our experiments delineate three distinct failure modes with different signatures. (1) \emph{Statistical starvation} ($N/C\lesssim 1.5$): means are estimated from ${\sim}1$ clip per class; errors compound across iterations (the batch-64 trajectory reversal of Sec.~\ref{ssec:main}); the fix is simply larger batches or restricted label spaces. (2) \emph{Prior mismatch} (long-tailed batches): gains attenuate smoothly and predictably, accuracy stays above zero-shot, and iterations remain stable---a bias, not a collapse. (3) \emph{Absent alignment} (TUT scenes): the prior $z^0$ carries no class signal, so there is nothing to refine; no batch size helps, and the supervised-probe gap (about $41\%$ on TUT vs.\ about $87\%$ on US8K) shows the deficit lies in the representation, upstream of any transductive machinery. Practitioners can diagnose all three without labels: (1) from $N/C$, (2) from the entropy of the aggregate posterior $\frac1N\sum_i z_i$, and (3) from near-uniform per-clip posteriors.

\section{Limitations and future work}
\label{sec:limitations}
Beyond the uniform-prior sensitivity of Sec.~\ref{ssec:longtail}, four limitations qualify our claims. (i) All results use one public checkpoint (\texttt{laion/clap-htsat-unfused}); the operating boundary's \emph{location} (${\sim}2.5$ samples per class) may shift for stronger ALMs \cite{elizalde2024msclap,ghosh2024reclap}, although the $N/C$ mechanism itself is checkpoint-agnostic. (ii) The mixture model assumes single-label clips; multi-label audio tagging (e.g., FSD50K \cite{fonseca2022fsd50k}) would require replacing the softmax E-step with per-class responsibilities. (iii) Transduction presupposes that a batch is available at once; streaming audio would need running estimates of the class means. (iv) We deliberately kept the objective minimal; the covariance and Laplacian terms of full TransCLIP \cite{zanella2024transclip}, batch-prior estimation via Sinkhorn-style constraints \cite{cuturi2013sinkhorn}, and composition with stronger prompt optimization \cite{entropyprompt2026,hanif2024palm} are all orthogonal extensions that our ablations suggest would stack.

\section{Conclusion}
\label{sec:conclusion}
We presented the first systematic transfer of TransCLIP-style transductive inference to audio--language models. A minimal text-anchored GMM-EM over the unlabeled test batch---no labels, no gradients, milliseconds of compute---lifts CLAP zero-shot accuracy by $+4.6$ to $+9.2$ points across ESC-50, UrbanSound8K, and VocalSound, composes with prompt-side entropy weighting up to $96.2\%$ on ESC-50, and obeys a simple deployment rule of ${\gtrsim}2.5$ samples per class per batch. Gains attenuate gracefully under long-tailed batches and vanish when zero-shot alignment is absent; both regimes are detectable without labels. We hope this establishes batch-level transduction as a default, near-free test-time tool for audio--language models.

\bibliographystyle{IEEEbib}
\bibliography{refs}

\end{document}